\documentclass[12pt,a4paper,english,aps,preprint]{revtex4}
\usepackage[english]{babel}
\usepackage{graphicx,color}
\usepackage{latexsym}
\usepackage{amsmath}
\usepackage{amsfonts}
\usepackage{amssymb}
\usepackage[latin1]{inputenc}
\usepackage{hyperref}

\newcommand{\tr}{\mathrm{Tr}}

\newcommand{\be}{\begin{equation}}
\newcommand{\ee}{\end{equation}}
\newcommand{\bea}{\begin{eqnarray}}
\newcommand{\eea}{\end{eqnarray}}

\begin{document}

\title{Glueballs Mass Spectrum in an Inflationary Braneworld Scenario}

\author{L. Barosi, F. A.
Brito}\email{lbarosi@ufcg.edu.br, fabrito@df.ufcg.edu.br}\affiliation{{Departamento de
F\'\i sica, Universidade Federal de Campina Grande, Caixa Postal
10071, 58109-970 Campina Grande, Para\'\i ba, Brazil}}

\author{H. S. Jesuino}\email{hjesuino@gmail.com}\affiliation{{Instituto de F\'\i sica, Universidade de Bras\'\i lia,
Caixa Postal 04455, 70919-970 Bras\'\i lia, Brazil}}

\author{Amilcar R.
Queiroz}\email{amilcarq@unb.br}\affiliation{{Instituto de F\'\i sica,
Universidade de Bras\'\i lia,
Caixa Postal 04455, 70919-970  Bras\'\i lia, Brazil}\\{International
Center
for Condensed Matter Physics (ICCMP), Universidade de Bras\'\i lia,
Caixa Postal 04667, Bras\'\i lia, DF, Brazil}}

\begin{abstract}
We address the issue of glueball masses in a holographic dual field theory on the boundary of an $AdS$ space deformed by
a four-dimensional cosmological constant. These glueballs are related to scalar and tensorial fluctuations of the bulk
fields on this space. In the Euclidean $AdS_4$ case the allowed masses 
are discretized and are related to distinct inflaton masses on a 3-brane with several states of inflation. We then obtain the 
e-folds number in terms of the glueball masses. In the last part we focus on the Lorentzian $dS_4$ case to focus on the
QCD equation of state in dual field theory.
\end{abstract}
\maketitle
\tableofcontents

\section{Introduction}

In the large $N$ limit with the t'Hooft coupling
$\lambda=g_{YM}^2N>>1$ the $D3$-brane solution of type IIB
supergravity leads to an $AdS_5\times S^5$ spacetime. According to
the AdS/CFT correspondence \cite{malda97, Gubser:1998bc, witten98} string theory on $AdS_5\times S^5$ corresponds to a
${\cal N}=4$ superconformal $SU(N)$ gauge theory in $d=4$
dimensions. This holographic correspondence is better understood by
working out in the supergravity approximation \cite{csaki98}. In the
regime of small curvature of the spacetime (i.e. as $g_{YM}^2N>>1$)
compared to the string and Planck scale, string/M-theory can be well
described by supergravity.

The one-to-one correspondence between supegravity on $AdS_5\times S^5$ and
the ${\cal N}=4$ superconformal $SU(N)$ gauge theory fields
is given as follows. The mass $m$ of $p$-form field on the AdS space
(bulk) has well defined relation
with the conformal dimension $\Delta$ of a $(4-p)$ form operator in the
dual conformal guge theory
on the boundary in the form
\begin{equation}
m^2=(\Delta-p)(\Delta+p-4)
\end{equation}
In the spectrum of type IIB supergravity on $AdS_5\times S^5$ there
are four singlets of $SO(6)$ which corresponds to glueballs
\cite{csaki98}. Among them there are the massless graviton
$g_{\mu\nu}$ (on the bulk) that couples to the stress-energy tensor
$T_{\mu\nu}$ (on the four-dimensional boundary) with conformal
dimension $\Delta=4$ and the complex massless scalar field (on the
bulk) whose real part is the dilaton that couples to the scalar
operators tr$F^2$ and tr$F\wedge F$ of the four-dimensional theory
(on the boundary) with conformal dimension $\Delta=4$. The other
bulk fields are 2 and 4-form fields that we shall not discuss here.

In this paper we shall concentrate in the first two bulk fields. We
shall consider a geometry that can be embedded viewed as a
$D3$-brane solution in the limit of large $N$. The $D3$-brane
geometry and their deformed versions \cite{csaki98} are usually used
as good dual of glueballs in three dimensional chromodynamics
($QED_3$). The gravitational solution we use is the solution of five
dimensional braneworld scenario \cite{Karch:2001cw, Karch:2000ct,Binetruy:1999hy,Bazeia:2007vx}. Here one can understand that we are
turning on only the gravitational sector of a type IIB supergravity.
The bulk fields here couple to fields with the same conformal
dimension on the boundary. This is because in our case the equation
of motion for the fluctuations of the metric is formally the same
equation for the dilaton field.

These boundary fields are related to glueballs $2^{++}$ and $0^{++}$ and
their masses $m^2_{2^{++}}=m^2_{0^{++}}$
are given by the bulk equations of motion for the gravitational
fluctuations  and dilaton fields in the
gravitational background considered in the braneworld scenario. We point
out that the this glueball spectra is
also related to the {\it inflaton spectra} and we discuss several
inflationary states.

\section{Scenario}

In this section we describe a scenario of inflationary braneworld. It will be
in this scenario that we will obtain a glueball spectrum by analysing its
holographic dual field. This holographic dual field will be identified with an
inflaton field.

The inflationary braneworld metric in $d+1$ dimension is given by
\begin{equation}\label{mod_defayet_metric}
    ds^2=\alpha'\left[\frac{U^2}{R^2_{0}}\left( edt^2+a_0(t)^2 d\vec{x}_{d-1}^2
\right) + \frac{R^2_{0}}{\left(U^2-C \right)}~dU^2\right],
\end{equation}
where
\begin{equation}
    U(r)=({e\gamma+\xi^2e^{\mu r}+\chi^2e^{-\mu r}})^{\frac{1}{2}},
\end{equation}
with $e=-1$, for Lorentzian signature, or $e=+1$, for Euclidean
signature, also $\mu=(1/3)(\kappa_5^4\sigma^2)^{1/2}$, and
\begin{eqnarray}
    \label{xi-chi}
    \xi^2= \frac{1}{2}({1-e\gamma-\sqrt{1-2e\gamma}}\,), \qquad
    \chi^2= \frac{1}{2}(1-e\gamma+\sqrt{1-2e\gamma}\,),
\end{eqnarray}
with
\begin{equation}
    \gamma = \frac{d-1}{2\sigma}~H^2,
\end{equation}
and also
\begin{equation}
   \label{eq: def.C}
    C=2 R^4_{0}~\gamma ,
\end{equation}
where $H$ is the Hubble parameter (or the cosmological constant) in
the Braneworld, and $T_{brane} = \sigma/\alpha '^{d/2}$ is the
Braneworld tension. Note that only $\alpha'$ is dimensional with
$[\alpha']=\textrm{length}^2$, and other parameters with
$[R_{0}]=[\sigma]=[H]=[x^\mu]=1$. The Hubble parameter appears in
the braneworld warp factor in terms of the cosmological constant as follows
\begin{eqnarray}
 a_0(\tau)=\exp(H_E \tau),\:\:\:\:  H_E^2\propto - e\Lambda>0,\: e=+1,\: \Lambda<0,\qquad \mbox{(the Euclidean $AdS_4$ case)},\\   
 a_0(t)=\exp(H t),\:\:\:\:  H^2\propto - e\Lambda>0,\: e=-1,\: \Lambda>0,\qquad \mbox{(the Lorentzian $dS_4$ case)}.
\end{eqnarray}
The Euclidean and Lorentzian times relate with each other as $\tau=\sqrt{-e}t$.

The above metric may be written in Poincar\'{e}-like coordinates,
which will be useful later. For that, we first rewrite the above
metric as
\begin{equation}
    ds^2=\alpha'\frac{U^2}{R^2_{0}}\left[\left( edt^2+a_0(t)^2
    d\vec{x}^2_{d-1}
\right) + \frac{R^4_{0}}{U^2\left(U^2-C \right)}~dU^2\right].
\end{equation}
Now, we consider
\begin{equation}
    dz=\pm\frac{R^2_{0}}{U\sqrt{U^2-C}}~dU \equiv
\frac{R^2_{0}}{\sqrt{C}}~\frac{dy}{y\sqrt{y^2-1}},
\end{equation}
We obtain the metric in coordinates Poincaré-like
\begin{equation}
\label{metricPoinc}
      ds^2 = \alpha'\frac{U^2(z)}{R^2_{0}}\left[\left(
edt^2 + a_0(t)^2d\vec{x}^2_{d-1}\right) + dz^2\right].
\end{equation}
With $y=\frac{U}{\sqrt{C}}$. Integrating $dz$, we obtain
\begin{equation}
    z=\pm \frac{R^2_{0}}{\sqrt{C}}\sec^{-1}(y) + \textrm{const.}.
\end{equation}
where the choice of the constant to be related with the position of
the brane. Since we are considering $U>0$, we choose $y>0$.
Furthermore the constant of integration  can be suitably chosen so
that
\begin{equation}
    U = \sqrt{C}\csc\left(\frac{\sqrt{C}}{R^2_{0}}~z\right),
\end{equation}
\begin{equation}
    U = -\sqrt{C}\csc\left(\frac{\sqrt{C}}{R^2_{0}}~z\right),
\end{equation}
where we have used the definition (\ref{eq: def.C}). We now substitute
this change of variable in the metric, and after the scaling transformation
\begin{equation}
    x^\mu\to \frac{R_0^2}{\sqrt{C}}~x^\mu ~~~~ \textrm{ and } ~~~ z\to
 \frac{R_0^2}{\sqrt{C}}~z,
\end{equation}
we obtain
\begin{equation}
\label{eq:metric}
    ds^2 =
(R^2_{0}\alpha^{'})\left[\frac{1}{\sin^2\left(z\right)}\left( ds_{FRW}^2+dz^{2}
\right)\right].
\end{equation}
The FRW metric is given now by
\begin{equation}
    ds^2_{FRW} = edt^{2}+a_0(t)^2 d\vec{x}_{d-1}^2,
\end{equation}
where, as in the previous coordinates, we also have
\begin{eqnarray}
 a_0(\tau)=\exp(H^E_{\textrm{eff}} \tau),\:\:\:\:  {(H^E_{\textrm{eff}})}^2\propto - e\Lambda>0,\: e=+1,\: \Lambda<0,\:\:\: \mbox{(the Euclidean $AdS_4$ case)},\\   
 a_0(t)=\exp(H_{\textrm{eff}} t),\:\:\:\:  (H_{\textrm{eff}})^2\propto - e\Lambda>0,\: e=-1,\: \Lambda>0,\:\:\: \mbox{(the Lorentzian $dS_4$ case)},
\end{eqnarray}
with
\begin{equation}
    H_{\textrm{eff}}=\sqrt{\frac{\sigma}{d-1}}.
\end{equation}

\begin{figure}[h]
  \begin{center}
         \includegraphics[width=0.65\textwidth]{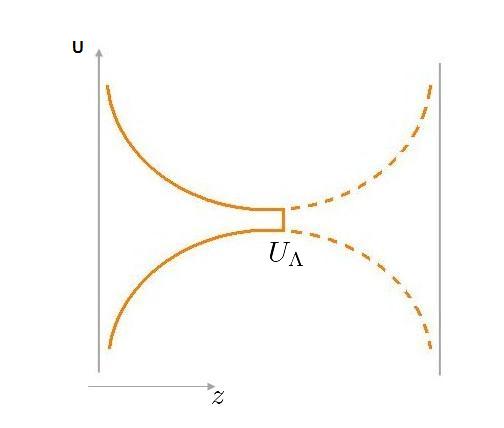}
          \end{center}
  \caption{The appearance of a natural infrared cut-off.}
  \label{fig1.5}
\end{figure}
Observe that the metric we are dealing with is similar to the metric in the work
of Karch and Randall \cite{Karch:2000ct}. The differences may be summarized as follows.
\begin{itemize}
    \item In Karch and Randall, they work with

       \subitem Lorentzian signature, i.e., $e=-1$,

       \subitem the cosmological constant in the braneworld can assume
the following values
            \subsubitem AdS, i.e., $\Lambda < 0$,
            \subsubitem dS, i.e., $\Lambda > 0$;

     \item In the present work,  we shall mainly consider the Euclidean $AdS_4$ case given by $e=+1$ with $\Lambda < 0$.
\end{itemize}

\section{Glueball and Inflaton Spectrum Equation}

In this section, we present arguments in favour of a description of the dual
of a $0^{++}$ glueball in the inflationary braneworld as an inflaton field.

Although the spacetime we are considering is not asymptotically AdS, it is a
conformally compact spacetime. Thus we may systematically use holographic
renormalization method. In this setting we may provide some arguments to the
claim that the dual to the glueball field is an inflaton field.

Metric (\ref{eq:metric}) may be seen as a simple deformation of the
AdS spacetime. One interesting feature of this metric is the
appearance of a natural infrared cut-off - see Fig.~\ref{fig1.5} -
depending on the Hubble parameter $H$.  Therefore, let's recap  the
usual set up in the usual AdS case.

In the gauge theory side, on top of a Minkowski $M_4$ spacetime (the boundary
of the AdS), we may consider the operator field $\hat{\mathcal{O}}\sim\tr(F^2)$.
As usual, correlation functions for this operator are obtained  by
computing the \emph{vev} of a generating
functional formed with the integral on $M_4$ of the coupling $\phi_0
\hat{\mathcal{O}}$, with $\phi_0$ seen as a auxiliary field. This
auxiliary field is a dual field that may be extended into the bulk of the AdS
spacetime as a massless dilaton field.

The holographic renormalization method prescribes that we might obtain the
above mentioned correlation functions by working with a renormalized on-shell
supergravity action. 

Now, instead of considering the Minkowski space above, we consider a $4d$ FRW
spacetime in inflationary regime, that is, $H=\dot{a}/a\equiv\textrm{constant}$.
As shown in section 2,
this space can be embedded into a deformed AdS spacetime. On top of this FRW
spacetime, we may consider the same operator  field
$\hat{\mathcal{O}}\sim\tr(F^2)$, and an auxiliary field $\phi_0$. Our claim is
that \emph{$\phi_0$ is an inflaton field}. We give arguments for this claim
below.

Let's consider the metric as given by (\ref{eq:metric}),
\begin{equation}
       \label{eq:metric2}
    \frac{ds^2}{R^{2}_{0}\alpha '} =\frac{1}{\sin(z)^2}\left(dz^2+edt^2+a_0^2
\delta_{ij}dx^i dx^j \right).
\end{equation}

Therefore, we may rewrite this metric by a proper choice
of a defining function. Recall that defining functions are defined
up to a scale transformation.

A defining  function for the spacetime we are considering is
$r(z)=\sin(z)$, as a defining function, i.e., a first order zero at
the boundary, $r(0)=0$, with $\partial_z r(0)\neq 0$, and $r(z)>0$,
for $0<z<\pi$. { Furthermore, in the neighbourhood of the boundary
$z=0$, we may write $\sin(z)\simeq z$, so that
\begin{equation}
    ds^2=\frac{1}{z^2}\left( dz^2 + g_{\mu \nu}(t) dx^\mu dx^\mu \right),
\end{equation}
where the $FRW$ metric is obviously smooth as $z\to 0$, so that we may write
\begin{equation}
    g_{\mu \nu}(t,z)=g^{(0)}_{\mu \nu}(t)+z~g^{(1)}_{\mu
\nu}(t)+z^2~g^{(2)}_{\mu \nu}(t)+... .
\end{equation}

As discussed in \cite{Skenderis:2002wp},
we may work in a fixed background. That means that we will work in
the case where the massless dilaton $\phi$ does not back-react into
the space-time. Therefore, we consider the gravitational side
action. We consider here a general $D=d+1$ dimension
spacetime, where $d$ refers to the braneworld space-time. Recall
that we are working in $D=4+1$.}

{ It is now well-known that the fluctuations around massless scalar and gravitational solutions obey formally the same
equation of motion of a scalar field coupled to gravity. Thus, to address our studies on glueballs we write down our
action below
\begin{equation}
\label{eq:action}
    S=\frac{1}{2}\int d^{d+1} x~\sqrt{eg}~\left({{\frac{1}{\kappa_{5}^2}(R + \Lambda_{bulk})}} + g^{M N}\partial_M \phi
\partial_N \phi+M^2 \phi^2\right).
\end{equation}
The massive scalar term will be removed later for consistence of the
conformal dimension of the relevant operator describing glueballs on
the dual four-dimensional field theory. The equation of motion for
$\phi$ is
\begin{eqnarray}
\label{eq:motion}\nonumber
e\left( \partial_0\partial^0 \phi + (d-1)H_{eff}\partial_0 \phi
\right) &+&
\frac{1}{a_0^2}\partial_i\partial^i \phi +\\
+ U^{1-d}\partial_z\left( U^{d-1} \partial^z \phi\right) &=& U^2 M^2 \phi.
\end{eqnarray}

}

We set from now on $\partial_i \phi=0$, since we want to discuss
homogeneous cosmology only. Furthermore, setting $\rho(z)\equiv
U(z)^{-2}$, we can look for solution of the form
\begin{equation}
    \phi(t,z)=\rho^{\frac{d-\Delta}{2}} \widetilde{\phi}(t,z).
\end{equation}
Substituting this ansatz in equation (\ref{eq:motion}), we obtain
\begin{eqnarray}
e\left(\partial_0\partial^0 \widetilde{\phi} +
(d-1)H_{eff}\partial_0
\widetilde{\phi} \right)    &+& \nonumber \\
\partial_z\partial^z \widetilde{\phi} + \rho' \rho^{-1}
\left(\frac{d+1}{2}-\Delta \right)~\partial_z \widetilde{\phi} & + & \nonumber
\\ \rho^{-1} \left( \rho'' -\frac{\Delta+1}{2}~\rho^{-1} (\rho')^2\right) ~
\left( \frac{d-\Delta}{2}\right) \widetilde{\phi} &=& M^2~\rho^{-1}
\widetilde{\phi}.
\end{eqnarray}
Now, since $\rho^{-1}\equiv U^2\equiv 1/\sin^2(z)$, we have
\begin{eqnarray}
    \rho' &=& 2\sqrt{\rho-~\rho^2} ~~~~ \textrm{and} \nonumber \\
    \rho'' &=& 2(1-2~\rho).
\end{eqnarray}
Therefore, the equation becomes
\begin{eqnarray}
     \label{eq:final}
    e\left(  \partial_0\partial^0 \widetilde{\phi} + (d-1)H_{eff}\partial_0
\widetilde{\phi} \right)    &+& \nonumber \\
\partial_z\partial^z \widetilde{\phi} + \rho' \rho^{-1}
\left(\frac{d+1}{2}-\Delta \right)~\partial_z \widetilde{\phi} & + & \nonumber
\\ (d-\Delta)~\left((\Delta-1)-\Delta\rho^{-1}\right) \widetilde{\phi}
&=& M^2~\rho^{-1}
\widetilde{\phi}.
\end{eqnarray}
Observe that, if we take $H_{eff}\to 0$, then $\rho\to 1/z^2$, and
the equation becomes
\begin{eqnarray}
\nonumber    e \partial_0\partial^0 \widetilde{\phi} +
\partial_z\partial^z
\widetilde{\phi} &+&
\\
+2\left(\Delta-\frac{d+1}{2} \right)~\frac{1}{z}\partial_z \widetilde{\phi} +
\left( \Delta(\Delta-d)- M^2\right)~z^2\widetilde{\phi}&=& 0.
\end{eqnarray}
This equation can be compared to similar equation obtained in the usual
$AdS_{d+1}$ case.

Now, we set $\widetilde{\phi}=T(t)\chi(z)$, so that we can separate equation
(\ref{eq:final}) into
\begin{eqnarray}
\label{EOMphit}
    e\left(\partial_0\partial^0 T + (d-1)H_{eff}\partial_0 T
\right)-m^2 T=0,\quad\quad\quad\quad\quad\quad\quad\quad\quad\quad\quad \\
\label{EOMphiz} -\partial_z\partial^z\chi-\left(\frac{d+1}{2}
-\Delta\right)\frac{\rho'\partial_z\chi}{\rho}+\left(
M^2-\Delta(\Delta-d) \right)\frac{\chi}{\rho}=\left(m^2-
~(\Delta-d)(\Delta-1) \right)\chi
\end{eqnarray}

In the equation for $\chi(z)$, we can consider the Ansatz
\begin{equation}
    \chi(z)=\rho^\alpha~\psi(z),
\end{equation}
so that this equation becomes
\begin{eqnarray}
    -\partial_z \partial^z \psi + \left(
\Delta-\frac{d+1}{2}-2\alpha\right)\frac{\rho'}{\rho}\partial_z \psi &+&
\nonumber \\
\left[ \left(
\Delta-\frac{d+1}{2}-(\alpha-1)\right)\frac{(\rho')^2}{\rho^2} +
\frac{\rho''}{\rho}\right.  &+& \nonumber \\
+ \left. \left( M^2-\Delta(\Delta-d)\right)\frac{1}{\rho}\right]
\psi &=& \nonumber \\ &=& \left( m^2-(\Delta-d)(\Delta-1) \right)
\psi.
\end{eqnarray}
This equation can be simplified into
\begin{eqnarray}
    -\partial_z \partial^z \psi+\left(
\Delta-\frac{d+1}{2}-2\alpha\right)\frac{\rho'}{\rho}\partial_z \psi &+&
\nonumber \\
    +\left(
M^2-\Delta(\Delta-d)\right)\frac{\psi}{\rho}
+2\alpha\left( 2\Delta-d-2\alpha\right) \frac{\psi}{\rho}&=& \nonumber \\
&=& \left( m^2-(\Delta-d)(\Delta-1)\right)\psi + \nonumber \\
&+&~2\alpha(2\Delta-(d+1)-2\alpha)~\psi
\end{eqnarray}
We may now set
\begin{equation}
    2\alpha=\Delta-\frac{d+1}{2}.
\end{equation}
We thus obtain
\begin{eqnarray}
    -\partial_z\partial^z \psi +\left(
M^2-\Delta(\Delta-d)\right)\frac{\psi}{\rho} &+& \nonumber \\
+\frac{1}{4}~\left(2\Delta-(d+1) \right)\left( 2\Delta-(d-1)\right)
\frac{\psi}{\rho} &=& \nonumber \\
&=& \left( m^2-(\Delta-d)(\Delta-1)\right)\psi + \nonumber \\
&+& \frac{1}{4}~\left( 2\Delta-(d+1)\right)^2~\psi.
\end{eqnarray}

Observe that if we set $d=4$, $M=0$ (massless dilaton), which implies
$\Delta=4$, then
\begin{eqnarray}
    e\left(  \partial_0\partial^0 T + 3H_{eff}\partial_0 T
\right) - m^2 T&=& 0, \\
-\partial_z\partial^z \psi + \frac{15}{4}\frac{1}{\sin^2{z}}~\psi
&=& \left( m^2 + \frac{9}{4}\right)~\psi.
\end{eqnarray}

In conclusion we have obtained an equation for the inflaton field with potential
given by  $V(T)=\frac{m^2}{2}~T^2$. This equation is similar to the equation
for conservation of energy in a FRW spacetime. Furthermore, the mass $m$ of the
inflaton field is quantized according to the equation for $\chi$. This equation
gives also the mass spectrum for the glueball. Recall that the final form for
the massless dilaton may be written as
\begin{equation}
    \phi(t,z)\equiv \rho^{\frac{d-\Delta}{2}} T(t) \chi(z)
=\rho^{\frac{d-\Delta}{2}} \rho^{\alpha} T(t) \psi(z)=
\left(\sin(z)\right)^{3/4}~T(t) \psi(z).
\end{equation}

\section{Glueball and Inflaton Spectrum}

In this section we obtain the spectrum of the glueball, that is, $m^2_n$, where
$n=1,2,3,...$. For that we have to solve the Schrödinger-like
equation
\begin{equation}
    -\chi(z)''+\frac{V_0}{\sin^2(z)}~
\chi(z) = \left(E+\frac{(d-1)^2}{4}~\right) \chi(z) \equiv \tilde{E}
\chi(z),
\end{equation}
with
\begin{eqnarray}
    V_0&=& \frac{d^2-1 }{4}, \\
    E\equiv m^2&=&\tilde{E}-\frac{(d-1)^2}{4}.
\end{eqnarray}

\subsection{Euclidean signature ($e=+1$)}

For this case, we set $y=\sin^{2}(z)$
\begin{equation}
    \chi(z)=y^\mu ~ \psi(y),
\end{equation}
which leads to the hypergeometric equation
\begin{equation}
    y(y-1)~\psi(y)'' +\left( (2\mu+\frac{1}{2} )-(1+2\mu)y \right) \psi(y)'
+\left( \frac{E}{4}-\mu^2\right) \psi(y) =0,
\end{equation}
with
\begin{equation}
    \mu=\frac{1}{4}( 1 + d),
\end{equation}
where $\mu$ was chosen to be positive. The solution for this hypergeometric
equation is
\begin{equation}
    \psi(y)=C_1~_2F_1(a,b;c;y)+ C_2 y^{1-c}~_2F_1(a+1-c,b+1-c;2-c;y),
\end{equation}
with
\begin{eqnarray}
    c&=& 1+2\mu, \\
    a &=& \mu + \frac{\sqrt{\tilde{E}}}{2}, \\
    b&=& \mu - \frac{\sqrt{\tilde{E}}}{2},
\end{eqnarray}
and $_2F_1(a,b;c;y)$ is a hypergeometric function. Now, using the
(Dirichlet) boundary conditions
\begin{equation}
    \chi(y=0)=0~\hspace{1cm}~\chi(y=1)=0,
\end{equation}
we have that $C_2=0$, and the solution
\begin{equation}
    \chi_n(y)=\mathcal{N}_n~ y^\mu~_2F_1(a,b;c;y),
\end{equation}
where $a=2\mu-n$, $b=-n$, with $n\in \mathbb{N}$, and $\mathcal{N}_n$ is a
normalization factor, so that
\begin{equation}
    \tilde{E}_n = 4~(\mu+n)^2.
\end{equation}
Therefore,
\begin{equation}
m^2_n = \left(
4(\mu+n)^2-\frac{(d-1)^2}{4}\right)=\frac{1}{4}~\left(
(1+d+4n)^2-(d-1)^2\right),
\end{equation}
for $n=0,1,2,...$. By relabelling $n\to \frac{n-1}{2}$, the mass spectrum
becomes
\begin{equation}
    m^2_n = n(n+d-1).
\end{equation}
In the case $d=4$, we obtain \cite{Karch:2001cw, Karch:2000ct}
\begin{equation}
    m^2_n = n(n+3),
\end{equation}
for $n=1,2,3,...$.

\subsection{Lorentzian signature ($e=-1$)}

In this case, it is well-know that there are only a finite number of bound
states for the potential. Furthermore there are scattered states, which is not
there in the previous $e=+1$ case. We shall be back to this case in the last
part of the paper.

\section{Slow-Roll and It's Consequences}

In this section, we analyse some consequences imposed by the
requirement of slow-roll condition. The slow-roll condition is
considered in order that the Universe goes through a nearly
exponential expansion during an Euclidean time $\tau\sim 1/H^E_{eff}$. This
condition is obtained by requiring that $\partial_0\partial^0
T_{n}(\tau)=0$, that is, the acceleration term in the equation
\begin{equation}\label{Inflaton}
    \left(  \partial_0\partial^0 T_n + (d-1)H^E_{eff}\partial_0 T_n
\right) - m_n^2 T_n= 0
\end{equation}
is negligible.  Since we shall focus on the transition between two particular
inflaton states we do not assume a priori a multi-inflaton cosmology. Thus our
slow-roll analysis is based on the equation for an inflaton field $T_n$ and the
induced Friedman equation \cite{Binetruy:1999hy,Bazeia:2007vx} given by its Euclidean form
\begin{eqnarray}\label{Friedman}
    {(H^E_{eff})}^2&=&-\frac{2}{3}\rho\left(1+\frac{\rho}{2\sigma}\right)=-\frac{2}{3}\rho,
\qquad \mbox{for heavy brane, $\sigma\gg1$}\nonumber\\
&=&-\frac{2}{3}\left(\frac12\dot{T_n}^2+V(T_n)\right)\simeq -\frac{2}{3}V(T_n).
\end{eqnarray}
Notice that the potential has the right sign, $V(T_n)<0$, in Euclidean time as we can check from equation (\ref{Inflaton}). Below we
study the slow-roll conditions for such potential.

This slow-roll condition can be translated into two conditions
for the potential $V(T_n)=-\frac12m_n^2T_n^2$ and its derivatives as
\begin{eqnarray}
    \epsilon\equiv \frac{1}{2}~\left( \frac{V'(T_n)}{V(T_n)}\right)^2&\ll& 1 \\
    \eta\equiv \frac{1}{d-1}~\frac{V''(T_n)}{V(T_n)} &\ll& 1.
\end{eqnarray}
Now, that is possible if the condition
\begin{equation}
\label{condition_slow-roll}
   - V(T_n)''\ll(d-1)^2H_{eff}^2~\hspace{0.2cm}~\textrm{i.e.} ~~~
\left(\frac{m}{(d-1)H_{eff}}\right)^2\ll 1.
\end{equation}
Thus, in this case, the Friedman equation (\ref{Friedman}) and the
inflaton equation (\ref{Inflaton}) lead to the well-known solution
in the the slow-roll regime \begin{eqnarray}
\label{solution-slow-roll regime-a}
a_0(\tau)&=&a_0\exp{[m_n C(T^0_n\tau+\frac12Bm_n\tau^2)]},\\
\label{solution-slow-roll regime-b}T_n(\tau)&=&T^0_n+Bm_n\tau,
\end{eqnarray}
where $B$ and $C$ are dimensionless constants.
The inflationary regime with an exponential
($H^E_{eff}\equiv\frac{\dot{a}_0}{a_0}= const.$) fast growth is valid
as long as the masses $m_n^2\ll1$. { This is also endowed with the fact that the Euclidean time $\tau\equiv\beta=1/T$, so that
at high temperature (in beginning of the inflation) this term is subleading. On the other hand, the inflation {\it never}
ends because the quadratic term in $a_0(\tau)$ starts to dominate for later Euclidean times.
So from now on, when we will mention `end of
inflation' we will mean the end of the exponential inflation.}

Let's now implement the condition for slow-roll inflation into the mass
spectrum of the fluctuation. Recalling that the inflaton and the fluctuation
have to have the same mass spectrum $m_n^2$. Therefore
\begin{equation}
    m_n^2 = n(n+d-1).
\end{equation}
We now replace this condition into the slow-roll condition
(\ref{condition_slow-roll}),
\begin{equation}
     \frac{m_n^2}{(d-1)^2{(H^E_{eff})}^2}=\frac{n(n+d-1)}{(d-1)^2}~\frac{1}{{(H^E_{eff})}^2}.
\end{equation}
we obtain
\begin{equation}
    \frac{m_n^2}{(d-1)^2H_E^2}\equiv \frac{n(n+d-1)}{(d-1)^{2}}\frac{1}{{(H^E_{eff})}^{2}} \ll
1,
\end{equation}
substituting $H^E_{eff} = \sqrt{\sigma/(d-1)}$, so that
\begin{equation}
    n(n+d-1)\ll (d-1)~\sigma.
\end{equation}
Now, since $n=1,2,3,...$, we have to solve the following system of inequalities
\begin{eqnarray}
    n&>&0, \\
    n^2+(d-1)n-(d-1)^2~{(H^E_{eff})}^{2}&\ll& 1.
\end{eqnarray}

\subsection{The transition between states of inflation and the e-folds number}
Let us first consider the e-folds for a transition in between the states $j\to k$, { where the
inflaton does not roll but rather suffer mass transitions $m_j\to m_k$ whereas its initial value
are $T^0_j$ and $T^0_k$}. We
one can estimate the time interval of such a transition as
\begin{equation}
\Delta
\tau\sim\frac{2\pi}{m_k-m_j}=\frac{2\pi}{\sqrt{k(k+3)}-\sqrt{j(j+3)}}.
\end{equation}
Recall that $m_n^2=n(n+3)$. As we have previously discussed the exponential
inflationary regime takes place for very small inflaton masses that
we assume to be the ground state ($j=1$) and the end of such an inflation
starts for a very large `final' mass, i.e., $k=n_f\gg1$, { which
of course implies $m_f\gg m_i$}. Thus, we have now the simpler
equation for the transition time inflation
\begin{equation}
\Delta \tau\sim \frac{2\pi}{n_f}.
\end{equation}
\begin{figure}[h]
  \begin{center}
         \includegraphics[width=0.65\textwidth]{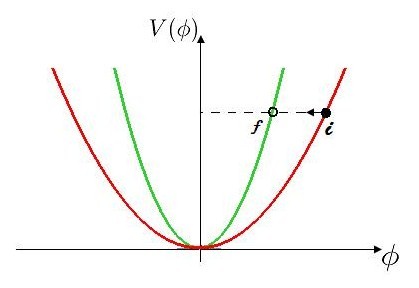}
          \end{center}
  \caption{Transition betwen masses $m_{i}$ $\rightarrow$ $m_{f}$. Notice that in the Euclidean case we should consider $-V$.}
  \label{fig.2}
\end{figure}

Now we are ready to estimate the number of e-folds as given by the
following equations
\begin{eqnarray}
N_{e}&=&\int_{\tau_{j}}^{\tau_{k}}{H^E_{eff}d\tau}\simeq H^E_{eff}\Delta \tau\nonumber\\
&=&\frac{2\pi}{n_f}~H^E_{eff} = \frac{2\pi}{m_f}H^E_{eff}.
\end{eqnarray}
One may find from the above relations the most useful relation as
being that gives the heaviest inflaton/glueball mass in terms of the
e-fold number and the Hubble constant
\begin{equation}
m_f=\frac{2\pi H^E_{eff}}{N_e}.
\end{equation}
{ Notice $({H^E_{eff}})^2\sim - V(T_n)$ during the inflation, with $V=-\frac12m^2_i{T^0_i}^2=-\frac12m^2_f{T^0_f}^2$. Now substituting $V$ in terms
of initial quantities we find
\begin{equation}
\frac{m_f}{m_i}\sim\frac{2\pi}{N_e}\frac{T_i^0}{m_{Pl}}.
\end{equation}
Note that for the usually assumed initial condition $T_i^0\sim m_{Pl}$ one finds that for $m_f\gg m_i$ one needs $N_e\to0$. Thus in the aforementioned transition no inflation really occurs.

Alternatively, one can also understand the previous analysis comparing two inflationary scenarios with distinct rolling inflaton solutions $T_i$  and $T_j$ given in Eq.~(\ref{solution-slow-roll regime-b}). Assuming they start and end with the conditions $T_i^0(\tau_0)=T_j^0(\tau_0)=T_0$ and $T_i^f(\tau^i_f)=T_j^f(\tau^j_f)=T^f$ one can find the final evolution times
\begin{equation}
\tau_f^i\sim \frac{T^f-T_0}{m_i},\qquad \tau_f^j\sim \frac{T^f-T_0}{m_j}.
\end{equation}
The respectively e-fold numbers can be found as follows
\begin{equation}
N_e^i=\int_{\tau_i=0}^{\tau_f^i}Hd\tau=H\tau_f^i\sim\frac{H(T^f-T_0)}{m_i},\qquad N_e^j=\int_{\tau_i=0}^{\tau_f^j}Hd\tau=H\tau_f^j\sim\frac{H(T^f-T_0)}{m_j},
\end{equation}
from what follows the ratio
\begin{equation}
\frac{N_e^i}{N_e^j}=\frac{m_j}{m_i}.
\end{equation}
Recall that $i<j$. Suppose $i=1$ and the e-folds number is that usually accepted in inflationary cosmology, say $N_e^1=60$,  then
\begin{equation}
{N_e^j}=60\frac{m_1}{m_j}.
\end{equation}
Notice that from this formula it is clear that for any other inflationary scenario other than $j=1$ the e-folds number $N_e^j$
decreases since $m_j>m_1$ for $j\neq1$. In the state with heaviest mass $m_j\to\infty$ of course  $N_e^j\to0$ and
the Universe never inflates. On another perspective one can think of that a Universe populated with large glueball masses cannot
inflate, but if it decays into lower glueball states it starts to inflate.

Thus, we should focus only on the slow-roll of the lowest state of inflation.  The inflaton is then slow-rolling with a determined state with mass $m_n$ that according to the previous discussions should be the one with smallest mass $m_1$ such that one achieves sufficient inflation. Before ending this subsection some interesting comments are in order. The theory we start with is a pure scalar theory coupled
to five-dimensional gravity given in Eq.~(\ref{eq:action}). For absence of dilaton masses ($M=0$) there is no chance of appearing inflation in $d=5$. However, in the
four-dimensional theory one has the inflaton fields describing glueballs which may produce sufficient inflation for sufficient small mass
(i.e., for the `easiest' four-dimensional scalar potential). Such a dimensional reduction was made by considering a deformed $AdS_5$ with time-independent fifth dimension. This certainly is not the case as one uses general Kaluza-Klein compactifications. As it was shown in \cite{gibbons,nunez} a theory with no inflation
in higher dimensions cannot generate lower dimensional theories with inflation by simple Kaluza-Klein reductions. The first exception was
shown in \cite{Townsend:2003fx} as one considers time-dependent internal coordinates on a manifold with negative cosmological constant.

\subsection{The slow-roll in $d=4$ versus entropy in $d=5$ dimensions}

In this subsection let us first discuss the e-folds number formula obtained in the slow-roll regime of the 
the four-dimensional Euclidean cosmology previously discussed. In this case the e-folds number reads 
\begin{eqnarray}
N_{e}&=&\int_{\tau_{j}}^{\tau_{k}}{H^E_{eff}d\tau}\simeq H^E_{eff}\Delta \tau=H^E_{eff}\Delta \beta=H^E_{eff}\left(\frac{1}{T_k}-\frac{1}{T_j}\right)
\nonumber\\
&=&H^E_{eff}\left(\frac{T_j-T_k}{T_jT_k}\right)\to dN_e=H^E_{eff}\frac{dT}{T^2},
\end{eqnarray}
where in the last step we have assumed the temperatures $T_i\sim T_k \sim T$ and $H^E_{eff}\sim const.$ Now 
we assume $dU=dT$ (from the equipartition theorem) and $dS=dU/T$ (from the thermodynamics second law) to write
\begin{eqnarray}
dN_e=\frac{H^E_{eff}}{T}\frac{dU}{T}=\frac{H^E_{eff}}{T}dS=\frac{H^E_{eff}}{T}Vds.
\end{eqnarray}
Notice that $s\equiv S/V$ is the entropy density and for relativistic particles goes like $s\sim T^3$. $V\sim a(\tau)^3$ is the comoving volume element. Conservation of
$S$ implies that $s\propto a(\tau)^{-3}$, such that $V\propto s^{-1}$. Also, since $S\equiv sV=const.$ implies that $a(\tau)\sim 1/T=\beta=\tau$, thus as a consequence $H^E_{eff}\equiv\frac{\dot{a}(\tau)}{a(\tau)}\sim T$. 

Now applying these considerations into our previous formula we find the interesting relationship between entropy density and e-folds number
\begin{eqnarray}
dN_e=\frac{ds}{s}\to N_e=\ln{s}.
\end{eqnarray}

In the following we deal with an interesting similar relationship between the e-folds number and dual black-hole entropy formulas in 
five dimensions.
This is so because the the slow-roll mechanism may be applied in both temporal and spatial coordinates in the suitable
dimension. The black-hole entropy formula we mean here stands for that one considered recently in Ref.~\cite{gubser} in
the context of equation of state of QCD via dual black-hole solutions in $d=5$ gravity coupled to a scalar field. We should emphasize
that we have no black holes in our set up but the deformed Euclidean $AdS$ has an apparent horizon that works as well as in the black
hole case.

In this last part we shall be considering the Lorentzian signature ($e=-1$). Notice that the equations of motion (\ref{EOMphit}) and (\ref{EOMphiz}) can be rewritten as
\begin{eqnarray}
\ddot{T}+3H\,\dot{T}+V_T=0, \qquad V(T)=\frac12m^2T^2,\\
\label{slow-roll-z}
\chi''+3\frac{U'}{U}\chi'+V_\chi=0, \qquad V(\chi)=\frac12m^2\chi^2,
\end{eqnarray}
where we have considered $\Delta=d=4$, $e=-1$ and $M=0$. Notice also that $U^2=1/\rho$ then follows that $2U'/U=-\rho'/\rho$. Analogously
to the slow-roll regime of $T(t)$ one finds the ``slow-varying'' $\chi(z)$ such that Eq.~(\ref{slow-roll-z}) is approximated by
\begin{eqnarray}
\label{slow-roll-z-2}
3\frac{U'}{U}\chi'+V_\chi=0.
\end{eqnarray}

Let us now focus on the Einstein equations. 
The relevant Einstein equations are given by the $0-0$ components
\begin{eqnarray}
\label{Einstein.1} -3
\left(\frac{\dot{a}_{0}(t)}{a_{0}(t)}\right)^{2} - 3
e\frac{U^{''}(z)}{U(z)} + \frac{\Lambda_{bulk}\alpha'e
U^{2}(z)}{2R_{0}^{2}} =
\kappa^{2}_{5}\left(-\frac{1}{2}\dot{\phi}^{~2} +
\frac{e}{2}\phi'^{~2} + \frac{\alpha'e
U^{2}(z)~M^{2}}{2R_{0}^{2}}\phi^{2}\right),
\end{eqnarray}
the $i-i$ components
\begin{eqnarray}
\label{Einstein.2.3.4}\nonumber -2e\frac{\ddot{a}_{0}(t)}{a_{0}(t)}
- e\left(\frac{\dot{a}_{0}(t)}{a_{0}(t)}\right)^{2} &-& 3
\frac{U^{''}(z)}{U(z)} +
\frac{\Lambda_{bulk}\alpha'U^{2}(z)}{2R_{0}^{~2}} = \\
&=& \kappa^{2}_{5}\left(\frac{1}{2 e}\dot{\phi}^{~2} +
\frac{1}{2}\phi'^{~2} +
\frac{\alpha'U^{2}(z)M^{2}}{2R_{0}^{2}}\phi^{2}\right),
\end{eqnarray}
and finally for the $z-z$ components
\begin{eqnarray}
\label{Einstein.5}\nonumber
-3e\left(\frac{\dot{a}_{0}(t)}{a_{0}(t)}\right)^{2} -
3e\frac{\ddot{a}_{0}(t)}{a_{0}(t)} -
6\left(\frac{U^{'}(z)}{U(z)}\right)^{2} &+&
\frac{\Lambda_{bulk}\alpha'U^{~2}(z)}{2R_{0}^{2}} =\\
&& \kappa^{2}_{5}\left(\frac{1}{2e}\dot{\phi}^{~2} -
\frac{1}{2}\phi'^{~2} +
\frac{\alpha'U^{~2}(z)~M^{2}}{2R_{0}^{2}}\phi^{2}\right),
\end{eqnarray}
where we have used the metric in the form (\ref{metricPoinc}) and the fact that
$\phi\equiv\phi(t,z)$, such that $\partial_{i}\phi =0\ (i=1,2,3)$.

There is a vacuum solution satisfying these equations for slow rolling/varying field, i.e. $\dot{\phi}=0,\phi'=0$, massless
`dilaton' $M=0$ and
$\Lambda_{bulk}$ term exponentially suppressed. Such a solution (for $e=-1$) is given by
\begin{eqnarray}
\label{exp-z-t}
a_0(t)= e^{\sqrt{\Lambda}t},\qquad U(z)=e^{-\sqrt{\Lambda}z}.
\end{eqnarray}

Let us turn to the slow rolling/varying field regime. We shall consider now the less stringent conditions $\dot{\phi}\simeq0,\phi'\simeq0$. Recall that $\phi$ describes fluctuations that can
be written as $\phi(t,z)=T(t)\chi(z)$. Thus $\dot{\phi}=\dot{T}(t)\chi(z)$ and $\phi'=T(t)\chi'(z)$. The
fields $T(t)$ and $\chi(z)$ are slow rolling/varying consistently if they are given by $T_n(t)=T^0_n-Bm_n t$ and
$\chi_n(z)=\chi^0_n+Bm_nz$, being $m_n$ glueball masses. The slow roll regime is ensured as long as $m_n\ll1$ such
that we also maintain the conditions  $\dot{\phi}\sim m_n\chi(z)\simeq0$ and $\phi'\sim m_n T(t)\simeq0$.

Now notice that from $z-z$ component one can readily find (for $e=-1$)
\begin{eqnarray}
3\frac{\ddot{a}_{0}(t)}{a_{0}(t)}
+3\left(\frac{\dot{a}_{0}(t)}{a_{0}(t)}\right)^{2} -
6\left(\frac{U^{'}(z)}{U(z)}\right)^{2} &=&
\kappa^{2}_{5}\left(-\frac{1}{2}\dot{\phi}^{~2} -\frac{1}{2}\phi'^{~2}\right)\nonumber\\
&=& \kappa^{2}_{5}B^2\left(-\frac12m_n^2\chi^{2}(z) -\frac12m_n^2 T^{2}(t)\right).
\end{eqnarray}
In the slow rolling/varying field regime one the first and second derivative terms have approximately the same value such
that one can write the above equation in a simplest form
\begin{eqnarray}
\left(\frac{\dot{a}_{0}(t)}{a_{0}(t)}\right)^{2} -
\left(\frac{U^{'}(z)}{U(z)}\right)^{2} =\frac{\kappa^{2}_{5}B^2}{6}\left(-\frac12m_n^2\chi^{2}(z) -\frac12m_n^2 T^{2}(t)\right).
\end{eqnarray}
This gives us the desired Friedman equation for the spatial component
\begin{eqnarray}
\left(\frac{U^{'}}{U}\right)^{2} =-\frac{\kappa^{2}_{5}B^2}{6}\left(-\frac12m_n^2\chi^{2}(z) -\frac12m_n^2 T^{2}(t)-\frac{6H^2}{\kappa^{2}_{5}B^2}\right)=-\frac{\kappa^{2}_{5}B^2}{6}V(\chi),
\end{eqnarray}
being 
\begin{eqnarray}
V(\chi)=-\frac12m_n^2\chi^{2}-\frac{6H^2}{\kappa^{2}_{5}B^2},
\end{eqnarray}
where we have made $m_n^2 T^{2}(t)$ negligible for $a(t)=\exp{(H t)}$, being $H\sim const.$. Notice this potential has 
the property of having a local maximum around $\chi=0$ just as those required in \cite{gubser}.

Now we are ready to compute the entropy density and the temperature in terms of this potential
\begin{eqnarray}
\ln{s}\equiv N_e=\int{dz \frac{U'}{U }}=-\int_{\chi_0}^{\chi_H}{{d\chi}\frac{V(\chi)}{V'(\chi)}},\nonumber\\
\label{tempT}
\ln{T}=\int_{\chi_0}^{\chi_H}{{d\chi}\left(\frac12\frac{V'(\chi)}{V(\chi)}-\frac13\frac{V(\chi)}{V'(\chi)}\right)},
\end{eqnarray}
where $\chi_H$ is $\chi$ at horizon.
We are now able to find equation of state by using such formulas. The sound speed can be readily found in the form
\begin{eqnarray}
c_s^2&\approx& \frac13-\frac12\frac{V'(\chi_H)^2}{V(\chi_H)^2}\\
&=&\frac13-\frac12\frac{(m_n^2\chi_H)^2}{(\frac12m_n^2\chi_H^{2}+\frac{6H^2}{\kappa^{2}_{5}B^2})^2}.
\end{eqnarray}
One can easily find $\chi_H$ in terms of the temperature from the temperature formula in (\ref{tempT}) for $\chi_H\to0$. In doing this gives $\chi_H\sim{T^{-3/m_n^2}}$. The behavior of the quadratic sound speed $c_s^2$(up to fourth order in $\chi_H$) as a function of the
temperature is depicted in Fig.~\ref{sound-speed}.
\begin{figure}[h]
  \begin{center}
         \includegraphics[width=0.40\textwidth]{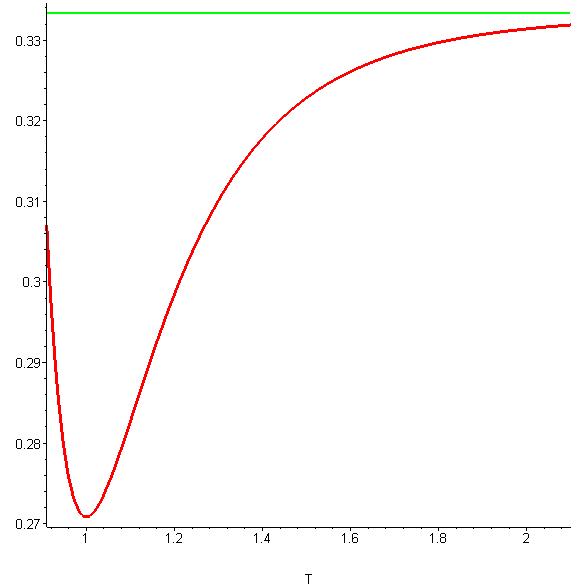}
          \end{center}
  \caption{The behavior of the sound speed as a function of the temperature (red) and the its asymptotic value 1/3 (green).}
  \label{sound-speed}
\end{figure}

}



{\bf Acknowledgments} 

The authors are supported in part by CNPq and PROCAD/PNPD-CAPES.





\end{document}